\documentclass[aps,prl,reprint,amsmath,amssymb,aps,]{revtex4-2}
\usepackage{tikz}
\usepackage{pgfplots}
\usepackage{graphicx}
\usepackage{dcolumn}
\usepackage{bm}
\usepackage{tikz}
\usepackage{tikz-feynman}
\usepackage{pgfplots}
\pgfplotsset{compat=1.18, width=8cm}
\usepgflibrary{arrows.meta}
\usepackage{listings}
\usepackage{color}
\usepackage{xcolor}
\usepackage{bbm}
\usepackage{blindtext}
\usepackage{hyperref}
\usepackage{mathtools}
\DeclareMathOperator{\Tr}{Tr}

\begin{document}
\preprint{APS/123-QED}
\title{Derivation of Fokker-Planck equation from Schr{\"o}dinger dynamics}
\author{Irfan Lone}
\thanks{{Correspondence.\and\\lone.irfan@candqrc.ca}}
\affiliation{Department of Physics and Astronomy, Howard University, Washington, DC 20059, USA.}
\affiliation{Canadian Quantum Research Center, 3002 32nd Ave Vernon, BC 204-3002 32, Canada.}
\affiliation{Department of Chemistry, Indian Institute of Technology Bombay, Powai, Mumbai 400076, India.}
\date{\today}
\begin{abstract}
The Fokker–Planck equation can be derived in a consistent manner through a microscopic approach based on a unified scheme of classical and quantum mechanics. Here we shall derive it through a purely quantum mechanical approach based on the reversible Schr{\"o}dinger dynamics. We also give a brief discussion of the path integral representation of the Fokker-Planck equation in light of our derivation. We conclude that, because of the use of the representation of eigenstates of the time-independent Hamiltonian in our derivation, the thermodynamical entropy in this case must correspond to a coarse-graining of the quantum entropy. 
\end{abstract}

\maketitle

Irreversibility is ubiquitous in nature. It is therefore an important question that how the apparently reversible fundamental laws of nature, for instance the Schr{\"o}dinger equation, give rise to this observed irreversibility in the physical world \cite{b1}. A commonly used approach to analyze this question is to ask how the laws governing the irreversible behavior, for instance those in classical statistical mechanics, arise from the underlying reversible laws operating at the fundamental level \cite{b1,b2,b3,b4,b5}. 

In this connection the Fokker-Planck equation has been derived in a consistent manner through a microscopic approach based on a unified scheme of classical and quantum mechanics \cite{b6}. The Fokker-Planck equation is a partial differential equation describing the time evolution of the probability density function of an observable characterizing some system of interest under the influence of some random force \cite{b7}. The equation has applications in information theory, graph theory, data science, finance, economics etc \cite{b8}. In addition, the formal analogy between the Fokker–Planck equation and the Schr{\"o}dinger equation has allowed the use of advanced operator techniques known from quantum mechanics in the solution of the former in a number of interesting cases \cite{b9,b10,b11,b12,b13}.

In this Letter we shall derive the Fokker-Planck equation through a purely quantum mechanical approach based on the reversible Schr{\"o}dinger dynamics.

Consider a quantum system described by a generic time-dependent Hamiltonian $H$ of the form
\begin{equation}
    H(t)=H_0+V(t),
\end{equation}
where $H_0$ is the time-independent part with known eigenstates and eigenvalues that satisfy
\begin{equation}
    H_0\lvert k\rangle=E_k\lvert k\rangle,
\end{equation}
and where the time-dependent part of the Hamiltonian, $V(t)$, induces transitions between the eigenstates of $H_0$. One can then write the wavefunction as 
\begin{equation}\label{e3}
    \lvert \psi(t)\rangle=\sum_ka_k(t)\lvert k\rangle
\end{equation}
The time evolution of $\lvert \psi(t)\rangle$ is governed by the Schr{\"o}dinger equation,
\begin{equation}\label{e4}
    i\hbar \frac{\partial \lvert \psi\rangle}{\partial t}= \hat{H}\lvert \psi\rangle
\end{equation}
Formally, the solution of Eq. (\ref{e4}) is written as
\begin{equation}\label{e5}
    \lvert \psi(t)\rangle=\mathcal{T}\exp{\left(-\frac{i}{\hbar}\int_0^tdt^{\prime}H(t^{\prime})\right)}\lvert \psi(0)\rangle,
\end{equation}
where $\mathcal{T}$ is the time ordering operator since the operators corresponding to the Hamiltonian at different times do not commute in general \cite{b13}. Restricting ourselves to time-independent Hamiltonians simplifies Eq. (\ref{e5}) to
\begin{equation}\label{e6}
    \lvert \psi(t)\rangle=\exp{\left(-\frac{i}{\hbar}Ht\right)}\lvert \psi(0)\rangle.
\end{equation}
Equation (\ref{e6}) is more concisely written as
\begin{equation}\label{e7}
    \lvert \psi(t)\rangle=U\lvert \psi(0)\rangle
\end{equation}
where $U$ is the unitary time evolution operator satisfying
\begin{equation}
    UU^{\dagger}=I
\end{equation}
and $I$ is the identity matrix. From Eqs. (\ref{e3}) and (\ref{e7}) one can obtain
\begin{equation}
    a_k(t)=\sum_lU_{kl}a_l(t)
\end{equation}
where $U_{kl}\equiv \langle k\lvert U\rvert l\rangle$, and where
\begin{equation}
    \sum_k\langle k|k\rangle=1
\end{equation}
has been used since the total probability for the system to be in any of the $k$ states must add up to unity. In terms of $U$, the occupation probability
\begin{equation}
    P_{k}(t) = \lvert a_k(t)\rvert^2
\end{equation} 
is expressed as
\begin{align}
    P_{k}(t)=\sum_{l,m}U_{kl}U_{km}^*a_l(t)a_m^*(t).
\end{align}
Separating the diagonal terms on the right-hand side, above equation becomes
\begin{equation}\label{e11}
    P_{k}(t)=\sum_lT_{kl}P_l(t)+\sigma_k(t),
\end{equation}
with $\sigma_k$ as the coherence terms
\begin{equation}
    \sigma_k(t)=\sum_{\substack{l,m\\(l\neq m)}}U_{kl}U_{km}^*a_l(t)a_m^*(t),
\end{equation}
where we have defined $T_{kl}=|U_{kl}|^2$ as the transition probability since
\begin{equation}
        T_{jk}\geq0
\end{equation}
and
\begin{equation}
    \sum_{j}T_{jk}=\sum_{k}T_{kj}=1
\end{equation}
Therefore, if the $\sigma_k$ terms could be neglected in Eq. (\ref{e11}), the time evolution of the occupation probability would be described by a classical Markovian process in which the transition probability for $k\rightarrow{l}$ in a time interval $\Delta t$, is given by $T_{kl}$. 

Provided the transition probability is differentiable with respect to time, one can then consider the transition probability per unit time $W_{kl}$, defined as
\begin{equation}
W_{kl}=\frac{T_{kl}-\delta_{kl}}{\Delta t},
\end{equation}
where $\delta_{kl}$ is the Kronecker delta \cite{b14}. $W_{kl}$ is also called the transition rate and $W_{kl}\delta t$ is the probability that the system jumps to state $k$ in an infinitesimal time interval $\delta t$, given that it is currently in state $l$. A straightforward calculation shows that equation (\ref{e11}) can then be written as
\begin{equation}
    P_k(t)=P_k(0)\nonumber\\
    +\sum_{l\neq k}(W_{kl}P_l-W_{lk}P_k)\Delta t+\sigma_k(t).
\end{equation}
The above equation is composed of two qualitatively different parts: one associated with a Markovian process, i.e., a classical-like diffusion, and the other, $\sigma_{k}$ associated with quantum coherence effects. It is this last term that preserves the unitary character of the evolution. For a purely classical process this term becomes negligible and the evolution is well approximated by
\begin{equation}\label{e16}
    \frac{\partial P_{k}}{\partial t}=\sum_{l\neq k}[W_{kl}P_{l}-W_{lk}P_{k}],
\end{equation}
The master equation (\ref{e16}) has an intuitive interpretation as a balance equation for the rate of change of the probability $P_k$ at site $k$ \cite{b14}. The first term on its right hand side describes the rate of increase of the probability at $k$ which is due to jumps from other states into $k$. The second term is the rate for the loss of probability due to jumps out of the state $k$. Equation (\ref{e16}) is sometimes also expressed as
\begin{equation}
    \frac{\partial P_{k}}{\partial t}=\sum_k\boldsymbol{M}_{kl}P_{l}(t),
\end{equation}
where the $\boldsymbol{M}_{kl}$ denote the Markov transition matrix elements describing the transition rates between sites $k$ and $l$ \cite{b15}. 
The above Markov process is said to satisfy detailed balance provided there exists a stationary distribution $\Pi_k$ such that for each pair of reversible transitions ($kl$):
\begin{equation}
    W_{kl}\Pi_l=W_{lk}\Pi_k.\nonumber
\end{equation}
In that case Eq. (\ref{e16}) may be written as,
\begin{equation}\label{e20}
    \frac{\partial P_{k}}{\partial t}=\frac{D}{2}\frac{\partial^2 P_{k}}{\partial k^2},
\end{equation}
which is the diffusion equation with the diffusion coefficient defined as
\begin{equation}
    D=2\sum_{l=1}^\infty W_{k,k+l}l^2.
\end{equation}
Thus the simplest form of the Fokker-Planck equation is the diffusion equation. If the initial condition is taken as
\begin{equation}
    P_k(0)=\delta(k),\nonumber
\end{equation}
where $\delta(k)$ is the Dirac delta function, the solution is given by
\begin{equation}
    P_k(t)=\frac{1}{\sqrt{2\pi t}}e^{-k^2/2t}.
\end{equation}
In arriving at Eq. (\ref{e20}) we have made use of the following discrete derivatives
\begin{equation}
    \frac{\partial P_{k}}{\partial t}=\frac{[P_k(t)-P_{k}(0)]}{\Delta t},
\end{equation}
and 
\begin{equation}
        \frac{\partial^2 P_{k}}{\partial k^2}=\frac{[P_{k+l}(t)+P_{k-l}(t)-2P_{k}(t)]}{l^2}.
\end{equation}
Equation (\ref{e20}) is thus a special case for the situation of a zero drift and a constant diffusion rate. For the drift $\mu$ and a diffusion coefficient that are not constant one has, on the other hand, the following generalized Fokker-Planck equation
\begin{equation}\label{e25}
    \frac{\partial P_{k}}{\partial t}=-\frac{\partial}{\partial l}[\mu_k(t)P_k(t)]+\frac{\partial^2}{\partial l^2}[D_k(t)P_k(t)]
\end{equation}
Finally, it is appropriate at this point to give a discussion of the path integral representation of the Fokker-Planck equation since the path integral formulation is an excellent starting point for the application of field theory based methods in the solution of this equation \cite{b16}. Before discussing the path integral representation of Fokker-Planck equation, however, it is instructive to give a brief account of the path integral formulation of quantum mechanics itself. 

The path integral formulation of quantum mechanics was developed by Richard Feynman \cite{b17}. This approach avoids the use of operators. Therefore, instead of finding the eigenfunctions of a Hamiltonian one here evaluates a functional integral that directly yields the propagator needed to determine the dynamics of a quantum system. To this end we proceed as follows.

From Eq. (\ref{e6}) one can write
\begin{equation}
    \langle k|\psi(t)\rangle=\int dk^{\prime}\langle k|\exp{\left(-\frac{i}{\hbar}Ht\right)}|k^{\prime}\rangle \langle k^{\prime}|\psi(0)\rangle
\end{equation}
Or
\begin{equation}
    \psi_k(t)=\int dk^{\prime} K(k,t,k^{\prime},0)\psi_{k^\prime}(0)
\end{equation}
with the propagator
\begin{equation}\label{e28}
    K(k,t,k^{\prime},0)=\langle k|\exp{\left(-\frac{i}{\hbar}Ht\right)}|k^{\prime}\rangle.
\end{equation}
This propagator is the central object of Feynman’s formulation of quantum mechanics \cite{b17}. It contains the complete information about the eigenenergies $E_n$ and the corresponding eigenstates $\lvert n\rangle$ of the quantum system. By making use of the completeness relation,
\begin{equation}
    \sum_n\lvert n\rangle \langle n\rvert=1,
\end{equation}
we obtain from Eq. (\ref{e28}),
\begin{equation}
    K(k,t,k^{\prime},0)=\sum_n\exp{\left(-\frac{i}{\hbar}E_nt\right)}\psi_n(k)\psi_n(k^{\prime})^*.
\end{equation}
In order to extract the information about the eigenenergies and eigenstates from the propagator, one introduces the retarded Green function
\begin{equation}
G_r(k,t,k^{\prime},0)=K(k,t,k^{\prime},0)\Theta(t)
\end{equation}
where $\Theta(t)$ is the Heaviside step function. A Fourier transformation gives the following spectral representation
\begin{equation}
\begin{gathered}
    G_r(k,k^{\prime},E)=-\frac{i}{\hbar}\int_0^{\infty}dt\exp{\left(\frac{i}{\hbar}Et\right)}G_r(t)\\
    =\sum_n\frac{\psi_n(k)\psi_n(k^{\prime})^*}{E-E_n+i\epsilon},
\end{gathered}
\end{equation}
where $\epsilon$ denotes an infinitely small positive quantity.

The main idea behind the path integral representation of the propagator is to decompose the time evolution over a finite time $t$ into $N$ slices of short time intervals
\begin{equation}
    \Delta t=\frac{t}{N}
\end{equation}
and eventually take the limit $N\rightarrow{\infty}$ \cite{b18}. We find
\begin{equation}\label{e34}
    \exp{\left(-\frac{i}{\hbar}Ht\right)}=\left[\exp{\left(-\frac{i}{\hbar}(T+V)\Delta t\right)}\right]^N.
\end{equation}
where $T$ and $V$ denote the operators for the kinetic and potential energy, respectively. Since we are assuming the Hamiltonian to be time-independent, we would now like to decompose the short-time propagator in Eq. (\ref{e34}) into a part dependent on the kinetic energy and another part on the potential energy. From an expansion of the Baker-Hausdorff formula we find
\begin{equation}
\begin{gathered}
    \exp{\left(-\frac{i}{\hbar}(T+V)\Delta t\right)}\approx \exp{\left (-\frac{i}{\hbar}T\Delta t\right)}\exp\left({-\frac{i}{\hbar}V\Delta t}\right)\\
    +\frac{1}{\hbar^2}[T,V](\Delta t)^2\nonumber
\end{gathered}
\end{equation}
where we have neglected terms of order $(\Delta t)^3$ and higher. Since we are interested in the limit $\Delta t\rightarrow{0}$, neglecting the contribution of the commutator gives the Trotter's formula
\begin{equation}
    \exp{\left(-\frac{i}{\hbar}(T+V)t\right)}=\lim_{N\rightarrow{\infty}}[U(\Delta t)]^N
\end{equation}
with the following short time evolution operator
\begin{equation}
    U(\Delta t)=\exp{\left(-\frac{i}{\hbar}T\Delta t\right)}\exp{\left(-\frac{i}{\hbar}V\Delta t\right)}.
\end{equation}
One now obtains for the propagator the following result
\begin{equation}
\begin{gathered}
    K(k_f,t,k_i,0)=\lim_{N\rightarrow{\infty}}\int_{-\infty}^{\infty}\left(\prod_{j=1}^{N-1}dk_j\right)\langle k_f|U(\Delta t)|k_{N-1}\rangle...\\
    \times \langle k_1|U(\Delta t)|k_i\rangle.
    \end{gathered}
\end{equation}
The propagator for a free particle is given by
\begin{equation}
    K(k_f,t,k_i,0)=\sqrt{\frac{m}{2\pi i\hbar t}}\exp{\left(\frac{i}{\hbar}\frac{m(k_f-k_i)^2}{2t} \right)}
\end{equation}
where $m$ is the mass of the particle \cite{b18}. Since the potential is diagonal in position representation, one obtains for the matrix element the result
\begin{equation}
\begin{gathered}
    \langle k_{j+1}|U(\Delta t)|k_j\rangle=\sqrt{\frac{m}{2\pi i \hbar \Delta t}}\\
    \times\exp{\left[\frac{i}{\hbar}\left(\frac{m}{2}\frac{(k_{j+1}-k_j)^2}{\Delta t}-V(k_j)\Delta t \right)\right]}.
\end{gathered}
\end{equation}
We thus finally arrive at the following expression for our propagator
\begin{equation}\label{e40}
\begin{gathered}
    K(k_f,t,k_i,0)=\lim_{N\rightarrow{\infty}}\frac{m}{2\pi i \hbar \Delta t}\int_{-\infty}^{\infty}\left(\prod_{j=1}^{N-1}dk_j \right)\\
    \times\exp{\left[\frac{i}{\hbar}\left(\frac{m}{2}\frac{k_{j+1}-k_j)^2}{\Delta t}-V(k_j)\Delta t \right) \right]}.
\end{gathered}
\end{equation}
Since the exponent in Eq. (\ref{e40}) contains a discretized version of the action
\begin{equation}\label{e41}
    \mathcal{A}[k]=\int_0^tds\left(\frac{m}{2}\dot{k}^2-V(k) \right),
\end{equation}
one can write this result in short hand notation as
\begin{equation}\label{e42}
    K(k_f,t,k_i,0)=\int\mathcal{D}k\exp{\left(\frac{i}{\hbar}\mathcal{A}[k]\right)}.
\end{equation}
The action $\mathcal{A}$ in Eq. (\ref{e41}) is a functional. The integral in Eq. (\ref{e42}) is thus a functional integral in which one integrates over all functions (paths) satisfying the boundary conditions $k(0) = k_i$ and $k(t) = k_f$. Hence the name path integral \cite{b18}.

The derivation of the path integral representation for the Fokker–Planck equation proceeds in a similar manner as above. We shall start by inserting a delta function and then integrate by parts as follows. Carrying out this procedure on Eq. (\ref{e25}) yields
\begin{equation}
    \begin{gathered}
        \frac{\partial P_{k^{\prime}}(t)}{\partial t}=-\frac{\partial}{\partial k^{\prime}}[\mu_{k^\prime}(t)P_{k^{\prime}}(t)]+\frac{\partial^2}{\partial k^{\prime^2}}[D_{k^{\prime}}(t)P_{k^{\prime}}(t)]\\
        =\int_{-\infty}^{\infty}dk\left[\left(\mu_k(t)\frac{\partial}{\partial k}+D_k(t)\frac{\partial^2}{\partial k^2}\right)\delta(k^{\prime}-k)\right]P_k(t).
    \end{gathered}
\end{equation}
where the $k$-derivatives here act only on the delta function and not on $P_k(t)$. Integrating over a time interval $\epsilon$ gives
\begin{equation}
\begin{gathered}
    P_{k^{\prime}}(t+\epsilon)=\\\nonumber
    \int_{-\infty}^{\infty}dk\left(\left(1+\epsilon\left[\mu_k(t)\frac{\partial}{\partial k}+D_k(t)\frac{\partial ^2}{\partial k^2}\right] \right)\delta(k^{\prime}-k)\right)P_k(t)\\\nonumber
    +\mathcal{O}(\epsilon^2).\nonumber
\end{gathered}
\end{equation}
Inserting the following Fourier integral for the delta function
\begin{equation}
    \int_{-i\infty}^{i\infty}\frac{d\tilde{k}}{2\pi i}e^{\tilde{k}(k-k^{\prime})}=\delta(k^{\prime}-k)
\end{equation}
gives
\begin{equation}
\begin{gathered}
    P_{k^{\prime}}(t+\epsilon)=\int_{-\infty}^{\infty}dk\int_{-i\infty}^{i\infty}\frac{d\tilde{k}}{2\pi i}\left(1+\epsilon \left[\tilde{k}\mu_k(t)+\tilde{k}^2D_k(t)\right]\right)\\
    \times e^{\tilde{k}(k-k^\prime)}P_k(t)+\mathcal{O}(\epsilon^2)\\
    =\int_{-\infty}^{\infty}dk\int_{-i\infty}^{i\infty}\frac{d\tilde{k}}{2\pi i}\\
    \times\exp{\left(\epsilon \left[-\tilde{k}\frac{(k^{\prime}-k)}{\epsilon}+\tilde{k}\mu_k(t)+\tilde{k}^2D_k(t) \right] \right)P_k(t)}\\
    +\mathcal{O}(\epsilon^2)
\end{gathered}
\end{equation}
The above expression expresses $P_{k^\prime}(t+\epsilon)$ as a functional of $P_k(t)$. However, upon iterating $(t^{\prime}-t)/\epsilon$ times and performing the limit $\epsilon \rightarrow{0}$ yields a path integral with action given by
\begin{equation}
    \mathcal{A}=\int dt\left[\tilde{k}\mu_k(t)+\tilde{k}^2D_k(t)-\tilde{k}\frac{\partial k}{\partial t}\right].
\end{equation}\\
where the variables $\tilde{k}$ conjugate to $k$ are dubbed the response variables, thus showing the formal equivalence of the two approaches, the Fokker–Planck equation its path integral formulation. 

In conclusion we have used the time-dependent Schr{\"o}dinger equation to arrive at the Fokker-Planck equation directly. While it has been demonstrated that it is possible to derive the later from the Schr{\"o}dinger dynamics we must mention here that the key assumption underlying our derivation is that the index $k$ corresponds to the eigenstates of the Hamiltonian $H_0$ and not to the eigenstates of the complete Hamiltonian $H(t)$. Obviously in the later case case the master equation (\ref{e16}) shall not hold. Since we have used the eigenstates of $H_0$, and not of $H(t)$, the thermodynamical entropy \cite{b19}, which must always increase,
\begin{equation}
    S\equiv-\sum_kP_k\ln{P_k}
\end{equation}
corresponds to a coarse-graining of the quantum entropy
\begin{equation}
\boldsymbol{S}\equiv -\Tr{(\rho\ln{\rho})}.
\end{equation}
which in this case must stay constant, where $\rho$ denotes the density matrix of the system \cite{b20}.


\begin{thebibliography}{apsrev4-2}






\bibitem{b1} E. A. Calzetta and B-L. B. Hu, {\color{blue}\textit{Nonequilibrium Quantum Field Theory}} (Cambridge University Press, 2008).
\bibitem{b2} G. Gallavotti and E. G. D. Cohen, {\color{blue}Phys. Rev. Lett. {\bf 74} 2694 (1995).}
\bibitem{b3} D. J. Evans, S. J. Searles, and L. Rondoni, {\color{blue}Phys. Rev. E {\bf 71} 056120 (2005).}
\bibitem{b4} U. Marini, B. Marconi, A. Puglisi, L. Rondoni, and A. Vulpiani, {\color{blue}Phys. Rep. {\bf 461} 111 (2008).}
\bibitem{b5} M. Colangeli and L. Rondonia, {\color{blue}Physica D {\bf 241} 681 (2012).}
\bibitem{b6} N. N. Bogolyubov Jr. and D. P. Sankovich, {\color{blue}Russian Math. Surveys {\bf 49} 19 (1994).}
\bibitem{b7} T. D. Frank, {\color{blue}\textit{Nonlinear Fokker-Planck equations: fundamentals and applications}} (Springer Science \& Business Media, 2005).
\bibitem{b8} W. Paul and J. Baschnagel, {\color{blue}\textit{Stochastic Processes: From Physics to Finance}} (Springer, 2013).
\bibitem{b9} T. D. Frank, {\color{blue}Phys. Rev. E {\bf 71}, 031106 (2005).}
\bibitem{b10} J. M. Heninger, D. Lippolis, and P. Cvitanovi{\'c}, {\color{blue}Commun. Nonlinear Sci. Numer. Simulat. {\bf 55}, 16 (2018).}
\bibitem{b11} F. M. Fern{\'a}ndez, {\color{blue}Phys. Scr. {\bf 80}, 065010 (2009).}
\bibitem{b12} T. Zhou, M-G Li, and S. Yan, {\color{blue}Phys. Rev. C {\bf 111}, 044001 (2025).}
\bibitem{b13} L. E. Ballentine, {\color{blue}\textit{Quantum Mechanics: A Modern Development}} (World Scientific Publishing Company, 2014).
\bibitem{b14} H. P. Breuer, E. M. Laine, J. Piilo, and B. Vacchini, {\color{blue}Rev. Mod. Phys. {\bf 88}, 1 (2016).}
\bibitem{b15} P. Rice, {\color{blue}\textit{An Introduction to Quantum Optics}} (IOP Publishing Ltd, 2020).
\bibitem{b16} J. Zinn-Justin, {\color{blue}\textit{Quantum field theory and critical phenomena.}} (Clarendon Press Oxford, 1996).
\bibitem{b17} R. P. Feynman, {\color{blue}Rev. Mod. Phys. {\bf 20}, 367 (1948).}
\bibitem{b18} H. Kleinert, {\color{blue}\textit{Path Integrals in Quantum Mechanics, Statistics and Polymer Physics}} (World Scientific 1995).
\bibitem{b19} R. Alicki and M. Fannes, {\color{blue}Lett. Math. Phys. {\bf 32} 75 (1994).}
\bibitem{b20} M. A. Nielsen and I. L. Chuang, {\color{blue}\textit{Quantum Computation and Quantum Information}} (Cambridge University Press, 2010).

\end{thebibliography}
\end{document}